\documentclass[12pt]{article}
\usepackage[dvips]{epsfig}
\usepackage{latexsym}
\usepackage{graphicx}
\usepackage{amssymb}
\usepackage{psfrag}
\usepackage[bf,footnotesize]{caption}	
\begin{document}
\setlength{\unitlength}{1mm}


\newcommand{\ket}[1] {\mbox{$ \vert #1 \rangle $}}
\newcommand{\bra}[1] {\mbox{$ \langle #1 \vert $}}
\def\vac{\ket{0}} 
\def\thermal{\ket{\beta}}
\def\bvac{\bra{0}}
\def\bthermal{\bra{\beta}}
\newcommand{\ave}[1] {\mbox{$ \langle #1 \rangle $}}
\newcommand{\avew}[1] {\mbox{$ \langle #1 \rangle $}_w}
\newcommand{\vacave}[1] {\mbox{$ \bvac #1 \vac $}}
\newcommand{\thermalave}[1] {\mbox{$ \bthermal #1 \thermal $}}
\newcommand{\scal}[2]{\mbox{$ \langle #1 \vert #2 \rangle $}}
\newcommand{\expect}[3] {\mbox{$ \bra{#1} #2 \ket{#3} $}}
\def\a{\hat{a}}\def\A{\hat{A}}
\def\b{\hat{b}}\def\B{\hat{B}}
\def\aa{\tilde{a}}\def\AA{\tilde{A}}
\def\bb{\tilde{b}}\def\BB{\tilde{B}}


\def\p{\prime}
\def\t{\tau}
\def\ga{\gamma}\def\Ga{\Gamma}
\def\om{\omega}\def\Om{\Omega}
\def\omp{\om^\p}\def\Omp{\Om^\p}
\def\la{\lambda}\def\lap{\lambda^\p}
\def\mup{\mu^\p}\def\lp{l^\p}
\def\kp{k^\p}\def\sig{\sigma} 
\def\al{\alpha}\def\alb{\bar\alpha}
\def\bt{\beta}\def\btb{\bar\beta}
\def\e{\epsilon}
\def\psip{\stackrel{.}{\psi}}\def\fp{\stackrel{.}{f}}
\def\VB{{\bar V}}\def\UB{{\bar U}}
\def\vb{{\bar v}}\def\ub{{\bar u}}
\def\lab{{\bar \la}}
\def\ffi{\varphi}
\def\scryp{{\cal J}^+}
\def\scrym{{\cal J}^-}
\def\scrypL{{\cal J}^+_L}
\def\scrymR{{\cal J}^-_R}
\def\scrypR{{\cal J}^+_R}
\def\scrymL{{\cal J}^-_L}

\def\div{\partial_V}
\def\divp{\partial_{V'}}
\def\divb{\partial_{\VB}}
\def\di{\partial}
\newcommand{\didi}[1]{\raise 0.1mm \hbox{$\stackrel{\leftrightarrow}{\di_{#1}}$}}


\def\TVV{T_{VV}}
\def\TVVI{T_{VV}^I}\def\TVVII{T_{VV}^{II}}
\def\TVVB{T_{{\VB}{\VB}}}
\def\Tvv{T_{vv}}
\def\Tvvb{T_{{\vb}{\vb}}}
\def\re{\mbox{Re}}
\def\im{\mbox{Im}}
\def\S{{\mathbf S}}
\def\T{{\mathbf T}}
\def\1{{\mathbf 1}}
\def\Ss{\mbox{$\hat S$}}

\def\disp{\displaystyle}
\def\bitem{\begin{itemize}}
\def\eitem{\end{itemize}}
\def\bes{\begin{description}}
\def\es{\end{description}}
\newcommand{\be} {\begin{equation}}
\newcommand{\ee} {\end{equation}}
\newcommand{\ba} {\begin{eqnarray}}
\newcommand{\ea} {\end{eqnarray}}

\def\cf{{\it cf}~}
\def\ie{{\it i.e.}~}
\def\etc{{\it etc}...}
\def\where{\mbox{where} \; \;}
\def\whereas{\mbox{whereas} \; \;}
\def\with{\mbox{with} \; \;}
\def\for{\mbox{for} \; \;}
\def\and{\mbox{and} \; \;}
\def\eg{{\it e.g.}~}

\def\nn{\nonumber \\}
\newcommand{\reff}[1]{Eq.(\ref{#1})}


\def\g{\tilde g}
\def\sgn{\mbox{sgn}}
\def\half{{1 \over 2}}
\newcommand{\inv}[1]{\frac{1}{#1}}

\def\inte{\int_{-\infty}^{+\infty}}
\def\into{\int_{0}^{\infty}}

\newcommand\Ie[1]{\inte \! d #1 \;}
\newcommand\Io[1]{\into \! d #1 \;}


\def\M{Minkowski}
\def\sg{\sqrt{-g}}
\def\sga{\sqrt{-\gamma}}
\def\gmn{g_{\mu\nu}}
\def\gnm{g^{\mu\nu}}
\def\gmn{g_{\mu\nu}}
\def\Rmn{R_{\mu\nu}}
\def\Rnm{R^{\mu\nu}}
\def\Gmn{G_{\mu\nu}}
\def\Gnm{G^{\mu\nu}}
\def\Tmn{T_{\mu\nu}}
\def\Tnm{T^{\mu\nu}}
\def\gnm{g^{\mu\nu}}
\def\Rij{R_{ij}}
\def\Rji{R^{ij}}
\def\Kij{K_{ij}}
\def\Kji{K^{ij}}
\def\gij{g_{ij}}
\def\gji{g^{ij}}
\def\piij{\pi_{ij}}
\def\piji{\pi^{ij}}
\def\gaij{\gamma_{ij}}
\def\gaji{\gamma^{ij}}
\def\H{{\cal H}}
\def\Hi{{\cal H}_i}
\def\Hii{{\cal H}^i}
\newcommand{\G}[3]{\Gamma^{#1}_{#2 \: #3}}
\def\A{{\cal A}_{BH}}

\overfullrule=0pt \def\sqr#1#2{{\vcenter{\vbox{\hrule height.#2pt
          \hbox{\vrule width.#2pt height#1pt \kern#1pt
           \vrule width.#2pt}
           \hrule height.#2pt}}}}
\def\lrpartial{\mathrel{\partial\kern-.75em\raise1.75ex\hbox{$\leftrightarrow$}}}

\newcounter{subequation}[equation] \makeatletter
\expandafter\let\expandafter\reset@font\csname reset@font\endcsname
\newenvironment{subeqnarray}
  {\arraycolsep1pt
    \def\@eqnnum\stepcounter##1{\stepcounter{subequation}{\reset@font\rm
      (\theequation\alph{subequation})}}\eqnarray}%
  {\endeqnarray\stepcounter{equation}}
\makeatother





\begin{flushright}
\end{flushright}
\vskip 1. truecm
\vskip 1. truecm
\centerline{\Large\bf{Uniformly accelerated mirrors}}
\vskip .2 truecm
\centerline{\large\bf{Part I : Mean fluxes}}
\vskip 1. truecm
\vskip 1. truecm

\centerline{{\bf N. Obadia}\footnote{e-mail: obadia@celfi.phys.univ-tours.fr}
 and {\bf
R. Parentani}\footnote{e-mail: parenta@celfi.phys.univ-tours.fr}}
\vskip 5 truemm\vskip 5 truemm
\centerline{Laboratoire de Math\'ematiques et Physique Th\'eorique,
CNRS-UMR 6083}
\centerline{Parc de Grandmont, 37200 Tours, France.}
\vskip 10 truemm

\vskip 1.5 truecm

\vskip 1.5 truecm
\centerline{{\bf Abstract }}
\vskip 3 truemm
\noindent
The Davies-Fulling model describes the scattering 
of a massless field by a moving mirror in $1+1$ dimensions. 
When the mirror travels under uniform acceleration,
one encounters severe problems which are due to the 
infinite blue shift effects associated with the horizons. 
On one hand, the Bogoliubov coefficients 
are ill-defined and the total energy emitted diverges. 
On the other hand, the instantaneous mean flux vanishes.
To obtained well-defined expressions
we introduce an alternative model based on an action principle.
The usefulness of this model is to allow 
to switch on and off the interaction at asymptotically large times. 
By an appropriate choice of the switching function,
we obtain analytical expressions for the scattering amplitudes 
and the fluxes 
emitted by the mirror. When the coupling is constant,
we recover the vanishing flux. However it is now 
followed by transients which inevitably become singular 
when the switching off is performed at late time. 
Our analysis reveals that the scattering amplitudes (and
the Bogoliubov coefficients)
should be seen as distributions and not as mere functions.
Moreover, our regularized amplitudes can be put in 
a one to one correspondence with the transition amplitudes of an accelerated detector,
thereby unifying the physics of uniformly accelerated systems.
In a forthcoming article, we shall use our scattering amplitudes
to analyze the quantum correlations amongst emitted particles 
which are also ill-defined in the Davies-Fulling model in the presence of horizons.

\vfill
\newpage


\section*{Introduction}

When considering the fluxes emitted by a uniformly accelerated mirror
described by the Davies-Fulling model,
one encounters a paradoxical situation: when working in the initial vacuum,
the local flux of energy vanishes whereas the Bogoliubov coefficients 
encoding pair creation do not.
Moreover,  $\ave{N_\om}$, 
the mean number of particles with frequency $\om$ emitted to $\scryp$
diverges \cite{DaviesFulling,Grove1,Recmir},
and so does the corresponding total energy 
$\ave{H} = \Io{\om} \om \ave{N_\om}$.

It should be pointed that similar properties are shared by all uniformly accelerated systems.
They are indeed found (in a slightly different form) in the case of 
a uniformly accelerated classical charge \cite{Boulware},
an accelerated two-level atom \cite{Unruh,Matsas,MaPa},
and for accelerated black holes \cite{Yi,MaPaBH}. 
The vanishing of the energy flux 
and the divergence of the total energy are both unavoidable when 
considering uniformly accelerated systems whose coupling to the radiation field is constant.
The vanishing of the flux follows from the facts that the orbits with uniform acceleration
are generated by the boost operator of the Lorentz group and that the Minkowski vacuum
is a null eigen-state of this operator. Hence stationarity is built in and this implies
a vanishing flux. (When applied to accelerated atoms this is known as
Grove's theorem \cite{Grove2,RSG,MPB,PhysRep}.) 
The divergence of the total energy emitted follows from the singular behavior
of the ``Rindler'' modes (i.e., the eigen-modes of the boost operator)
on the horizons associated to a uniformly accelerated trajectory,
see Appendix C of \cite{MaPa} for a general proof. 

In order to obtain regular expressions for the flux and the energy emitted, 
one needs either to regularize the trajectory by decreasing the acceleration
at asymptotically large times, or, what we shall do,  switch off the coupling
between the accelerated system (here the mirror) and the radiation field.
This procedure was already applied to an accelerated two-level atom in  \cite{MaPa}
and regular expressions have been found for the local flux and the total energy emitted. 
In the present paper, it is our intention to apply a similar treatment to an accelerated mirror. 
However, one immediately encounters a problem :
this analysis cannot be performed 
within the framework of the original Davies-Fulling model
because, by construction, the reflection on the mirror is total.
Therefore, we first introduce an alternative model 
based on an action principle which, on one hand, 
reproduces the results of the Davies-Fulling model 
when the coupling between the mirror and the radiation is constant and, on the other hand, 
allows to switch on and off the interactions.

When using this model to describe the scattering by an accelerated mirror,
we obtain regular expressions for the transition amplitudes in the place
of the singular Bogoliubov coefficients obtained with the Davies-Fulling model.
Both local quantities, such as the mean flux $\ave{\TVV}$, and global ones, 
such as the mean energy $\ave{H}$ and density of particles $\ave{N_\om}$, 
are now well defined.
Moreover, as long as the coupling is constant, we recover the
fact that the energy flux vanishes. However, it is now preceded and followed 
by transient effects associated with the switching on and off. 
Because of the ever increasing Doppler effect associated with constant acceleration, 
these effects are exponentially blue-shifted (in terms of the proper time of the mirror).
Therefore, in order to get a finite energy, the rate of switching off the interaction
must be faster than the growth of the Doppler effect 
(a condition also found in \cite{MaPa}). 
If this condition is not fulfilled, the mirror emits an infinite energy,
thereby recovering ill-defined results as those obtained with the Davies-Fulling model.   

To further clarify the physics into play in the scattering by a mirror of acceleration $a$,
we compute the transition amplitudes governing pair creation in a ``mixed'' representation.
By mixed we mean that in each pair, one quantum is characterized by $\om$, a Minkowski frequency,
whereas its partner is characterized by $\la$, a Rindler frequency, the eigenvalue of the energy  
with respect to the proper time of the mirror.
This representation is useful for the following reasons.
First, when $\la = \Delta M$, the scattering amplitudes closely correspond to the
transition amplitudes governing the absorption and emission 
of photons by an accelerated two-level atom
whose energy gap is $ \Delta M$. It is quite nice to see how,
for each Minkowski frequency $\om$, the ``exciton'' of the atom 
is replaced by the emission of the partner of the Minkowski quantum.
Secondly, the range of Minkowski frequencies $\om$ which participate to the processes
when the interactions last a proper time lapse of $T$ is finite and grows with $a e^{aT}$.  
Hence the physics into play is a succession of pair creation acts with Doppler shifted 
Minkowski frequencies. 
However, this cannot be seen
by considering only the expectation value of the flux, 
i.e. the one-point function of the stress tensor,
because these successive individual effects 
perfectly interfere with each order since, in the absence of recoils,
the orbit is generated by a Killing field. 
In order to reveal these local acts, one should either 
consider the correlations amongst emitted quanta by computing the
two-point function \cite{CarlitzWilley} of the energy flux or enlarge the dynamics
so as to take into account the recoil effects. In a next article\cite{OPa3}
we analyze these quantum correlations, and in a forthcoming paper we shall analyze 
recoil effects.

\section{The Davies-Fulling model and its extensions}

In this section, we first analyze 
the case of asymptotically inertial trajectories (Sec. $1.1$).
We introduce notations which also apply
to the cases of trajectories
which enter or leave space-time through null infinities (Sec. $1.2$), 
cases which turn out to be more delicate to analyze.
In Sec. $1.3$ we introduce the self-interacting model.
Sec. $1.4$ is devoted to the study of energy fluxes.

\subsection{Asymptotically inertial trajectories}

In the Davies-Fulling model \cite{DaviesFulling},
one studies the scattering of a massless scalar field 
induced by imposing a Dirichlet boundary condition 
along a time-like trajectory in $1+1$ dimensions.
The evolution of the field operator is governed 
by the d'Alembert equation 
\ba \label{d'Alembert}
(\partial_t^2 -\partial_z^2 )  
\Phi(t,z) 
= 0  \ ,
\ea
together with the reflection condition
along the trajectory of the mirror $z = z_{cl}(t)$ 
\ba \label{Dirichlet}
\Phi(t,z_{cl}(t))=0 \ .
\ea
Since we work in $1+1$ dimensions 
with a massless field,
it is particularly useful to work in the light-like coordinates 
$U, V = t \mp z$.
Then, \reff{d'Alembert} becomes 
$\di_U \di_V \Phi(U,V) = 0$,
and its general solution 
is a sum of a function of $U$ plus a function of $V$.
Since the trajectory of the mirror is given once
for all, the recoil effects induced by the scattering of quanta
are neglected. 

In this sub-section, in addition to the condition that 
the mirror trajectory be time-like, we consider only
asymptotically inertial trajectories,
\ie trajectories which originate from the time-like past infinity
$i^-$ and which end in $i^+$ (see Fig. $1$).
Then, since the reflection is total,
the configurations emerging from ${\scrymR}$, the right part of ${\scrym}$,
are completely decoupled from those emerging from ${\scrymL}$.
\begin{figure}[ht] 
\epsfxsize=6.5cm
\psfrag{scrypU}{${\scrypR}$}
\psfrag{scrypV}{${\scrypL}$}
\psfrag{scrymU}{${\scrymR}$}
\psfrag{scrymV}{${\scrymL}$}
\psfrag{im}{$i^-$}
\psfrag{ip}{$i^+$}
\psfrag{U}{$U$}
\psfrag{V}{$V$}
\centerline{{\epsfbox{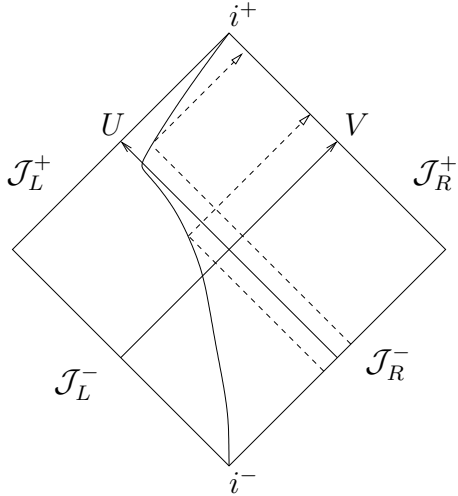}}}
\caption{In this Penrose diagram, 
the solid line is a time-like trajectory going from $i^-$ to $i^+$.
The dashed lines represent incoming $V$ configurations 
which give rise to the production of 
a pair of outgoing $U$ quanta
(right movers).}
\end{figure}
Therefore, one can analyze 
what happens on each side separately. 
On the right hand side, all left movers
are scattered into right movers and sent toward $\scrypR$.
In second quantization, when the trajectory is not inertial,
this leads to the production of pairs formed with two right movers.
Similarly, on the left hand side, one studies 
the scattering from $\scrymL$ to $\scrypL$.
Since the expressions governing the scattering
on the left are obtained from those on the right
by exchanging $U$ and $V$,
we will restrict ourselves to the analysis of the  
scattering from $\scrymR$ to $\scrypR$.

To analyze the scattering in second quantization,
one first needs to identify the $in$-basis of modes which are defined 
before the scattering occurs.
On $\scrymR$, the usual eigenmodes of Minkowski energy $i\di_t = \om > 0$ 
can be used since 
$\scrymR$ is a Cauchy surface for the
left movers when the mirror emerges from $i^-$
(this is no longer the case for trajectories which emerge from $\scrymR$, see Sec. $1.2$).
On $\scrymR$ the Minkowski modes are given by
\ba \label{inmodes}
\ffi_\om^{V,in}(V,U=-\infty) 
= \frac{e^{-i\om V}}{\sqrt{4\pi\om}} 
\; ,
\ea
where the upper index $V,in$ means that the mode 
is left-moving and defined on the initial Cauchy surface $\scrym$.
We have introduced the index $V$
in order to be able to deal with partially reflecting mirrors for which
left and right-movers should be simultaneously considered.
The norm of $\ffi_\om^{V,in}$ is determined by the usual
Klein-Gordon scalar product. 
On $\scrymR$, one has
\ba \label{normalization}
(\ffi_\om^{V,in} , \ffi_{\omp}^{V,in} )
&\equiv& \Ie{V} \ffi_\om^{V,in \: *} \: i \didi{V} \: \ffi_{\omp}^{V,in}
= \delta(\om - \omp) \ , 
\ea
where $f \: \didi{V} \: g = f \partial_V g - g \partial_V f$.

For finite values of $U$,
on the right hand side of the mirror, 
\ie for $V \geq V_{cl}(U)$,
the $in$-mode $\ffi^{V,in}_\om$,  
the solution of \reff{Dirichlet} which has \reff{inmodes} as initial 
Cauchy data, is
\ba \label{scatVU}
\ffi^{V,in}_\om (V,U) 
= 
\frac{e^{-i\om V}}{\sqrt{4\pi\om}}
- \frac{e^{-i\om V_{cl}(U)}}{\sqrt{4\pi\om}} \ .
\ea
To analyze its final frequency content,
it should be Fourier decomposed on $\scrypR$.
In total analogy with what we have on $\scrymR$, on  $\scrypR$
the positive frequency modes are
\ba \label{outmodes}
\ffi_{\omp}^{U,out}(U,V=+\infty) 
= \frac{e^{-i\omp U}}{\sqrt{4\pi\omp}} \ .
\ea
Since they are complete, on $\scrypR$,  the $in$-modes can be written as 
\ba
\label{scatUV}
\ffi^{V,in}_\om (U,V=+\infty) 
&=& \Io{\omp} \left(
\al^{UV \: *}_{\omp \om} \ffi^{U,out}_{\omp} 
- \bt^{UV \: *}_{\omp \om} \ffi^{U,out \: *}_{\omp}
\right) \ .
\ea
When evaluated on $\scrypR$, the overlaps are given by
\ba
\al_{\omp \om}^{UV \: *} \label{aUV}
&\equiv& ( \ffi_{\omp}^{U , out} , \ffi_{\om}^{V , in} )
= -2 \Ie{U} \frac{e^{i\omp U}}{\sqrt{4\pi / \omp}} 
\frac{e^{-i\om V_{cl}(U)}}{\sqrt{4\pi \om}} \ , \\
\bt_{\omp \om}^{UV \: *} \label{bUV}
&\equiv& ( \ffi_{\omp}^{U , out \: *} , \ffi_{\om}^{V , in} )
= 2 \Ie{U} \frac{e^{-i\omp U}}{\sqrt{4\pi / \omp }} 
\frac{e^{-i\om V_{cl}(U)}}{\sqrt{4\pi \om}} \ .
\ea

To interpret the scattering in terms of particle creation, 
one should decompose the Heisenberg field operator
$\Phi$ both in the $in$ and $out$-bases.
When working with a complex field, 
annihilation $in$-operators of particles and anti-particles are defined by
\ba
a_{\om}^{V, in} 
= \left( \varphi_\om^{V, in} , \Phi \right) \; , \;
b_{\om}^{V, in} 
= \left( \varphi_\om^{V, in} , \Phi^\dagger \right) \ .
\ea
Because the  $in$-modes $\ffi_\om^{V,in}$ form an orthogonal and complete basis,
these operators satisfy the canonical commutators when 
the field operator satisfies the equal-time commutation relation 
$[\Phi(t,z),\di_t \Phi^\dagger(t,z')] = i\delta(z-z')$.
The $in$-vacuum state is defined, as usual, by 
$a_{\om}^{V, in} \vac =  b_{\om}^{V, in} \vac = 0$.
Similarly, the $out$-operators are defined 
with the $out$-modes $\varphi_{\om}^{U, out}$.

Since we are dealing with a linear theory without sources,
the overlaps $\al$ and $\bt$ of Eqs.$(\ref{aUV})$ and $(\ref{bUV})$
define the Bogoliubov coefficients relating the initial and final operators $a_{\om}$,$b_{\om}$.
Therefore, these overlaps determine the expectation values (as well as the non-diagonal
matrix elements) of all operators built with $\Phi$.
For instance, when the initial state is vacuum,
the mean number of right-moving particles of energy $\omp$
received on $\scrypR$ is
\ba \label{NDF}
\ave{N^U_{\omp}}
\equiv {}_{in}\bvac a_{\omp}^{U, out \: \dagger} a_{\omp}^{U, out} \vac_{in}
= \Io{\om} \left| \bt_{\omp \om}^{UV} \right|^2 \ .
\ea

\subsection{Non-asymptotically inertial trajectories}

When the trajectory does not end on $i^+$ (or does not begin from  $i^-$),
the strict decoupling between left and right movers is no longer realized
even when the reflection on the mirror is total.
Consider for instance the trajectory
\ba \label{trajCW}
V_{cl}(U)=-\kappa^{-1} e^{-\kappa U}.
\ea
The mirror goes from $i^-$ to $V=0$ on $\scrypL$.
\begin{figure}[ht] 
\epsfxsize=6.5cm
\psfrag{scrypU}{${\scrypR}$}
\psfrag{scrypV}{${\scrypL}$}
\psfrag{scrymU}{${\scrymR}$}
\psfrag{scrymV}{${\scrymL}$}
\psfrag{im}{$i^-$}
\psfrag{ip}{$i^+$}
\psfrag{U}{$U$}
\psfrag{V}{$V$}
\centerline{{\epsfbox{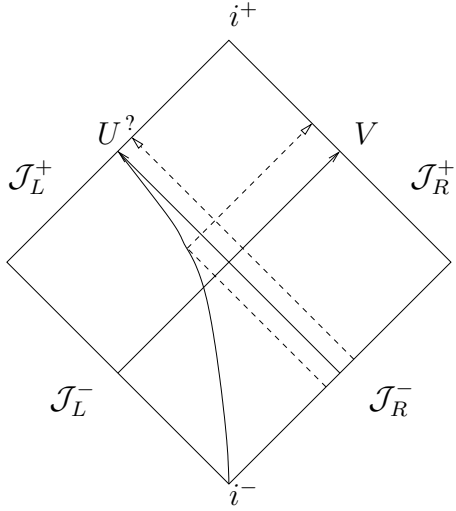}}}
\caption{In this Penrose diagram we represent the trajectory 
defined in \reff{trajCW}. This trajectory 
has been often considered (see \eg \cite{CarlitzWilley})
because of its analogy with the Hawking effect.
In this case, $\scrymR$ is a Cauchy surface whereas $\scrypR$ is not.
The portion of $\scrypL$ with $V>0$ plays the role 
of the future horizon of the black hole.
The dashed lines are incoming left-movers.
One sees that for $V<0$, the quanta are reflected, giving rise 
to right-movers for all values of $U$. On the contrary, for $V>0$,
the incoming quanta do not reach the trajectory and end up in $\scrypL$.
The question mark is there to raise the reader's attention
on the issue of the choice of the appropriate basis of out-modes 
to decompose the field configurations when the mirror crosses $\scrypL$.}
\end{figure}
Depending on the sign of $V$, 
left-movers emerging from $\scrymR$ are either 
reflected into right-movers for $V<0$
or end as left-movers on $\scrypL$ for $V>0$ (see Fig. 2).
Thus on $\scryp$, the image of $\ffi_\om^{V,in}$ of \reff{inmodes} 
now contains two pieces
\ba 
\ffi_\om^{V,in} (U,V) = \Theta(V) \frac{e^{-i\om V}}{\sqrt{4\pi\om}} 
- \frac{e^{i\om \kappa e^{-\kappa U} }}{\sqrt{4\pi\om}}  \ .
\ea
This mode is singular at $V=0$ where the mirror hits  $\scrypL$.
On the other side of the  mirror, the $U$ in-modes
emerging from $\scrymL$ are fully reflected into left-movers 
but they are also singular on $\scrypL$.
Hence both sets of $in$-modes are singular on $\scryp$, at $V=0$.

This raises an interesting question:
Given that the mirror trajectory ends on $\scrypL$,
which is part of the final Cauchy surface, 
what is the appropriate set of $out$-modes 
to describe the scattered field configurations ? 

The procedure we shall follow is to decouple asymptotically the radiation field from the mirror,
\ie to make the mirror asymptotically transparent. 
In this case, the free Minkowski modes $e^{-i\omp V}/\sqrt{4\pi\omp}$ and
$e^{-i\omp U}/\sqrt{4\pi\omp}$ 
still form a complete $out$-basis.
We shall not adopt the other choice which consists in working with
$out$-modes defined on either side of the mirror on $\scrypL$. 
These modes are so singular that their (Minkowski) energy content is not defined.
Nevertheless, when working in a state specified on $\scrym$, the
expectation values of local operators  whose
support is $V \neq 0$ is independent of the $out$-basis one chooses.
The $out$-basis is necessary only for computing global
quantities such as the total energy $\ave{H} = \Io{\om} \om \ave{N_\om}$.

When adopting the asymptotic decoupling hypothesis,
the image of $ \ffi_\om^{V,in}$ 
on $\scrypR \cup \{\scrypL(V>0)\}$ can be decomposed as
\ba \label{decomp}
\ffi_\om^{V,in}
= \Io{\omp} \Big( 
\al_{\omp \om}^{jV \: *} \; \ffi_{\omp}^{j,out} 
- \bt_{\omp \om}^{jV \: *} \; \ffi_{\omp}^{j,out \: *} 
\Big) \ ,
\ea 
where $\ffi_{\omp}^{j,out}$ are the usual Minkowski modes, as in \reff{scatUV}.
The Bogoliubov coefficients $\al,\bt$
are now $2 \times 2$ matrices. 
The discrete index $j$ stands for $U,V$ 
and is summed over when repeated.
These coefficients are still given by the Klein-Gordon scalar product
as in Eqs.(\ref{aUV}) and (\ref{bUV}),
since the out-basis is composed of usual Minkowski modes.
When the trajectory is asymptotically inertial, 
we recover what happens in the right hand side of the mirror for 
$\al^{ij}=\al^{UV},\bt^{ij}=\bt^{UV}$,
and on the left hand side for $\al^{ij}=\al^{VU},\bt^{ij}=\bt^{VU}$.
In addition, one has $\al^{VV}=\bt^{VV}=\al^{UU}=\bt^{UU}=0$.
Finally, we mention that
a similar decomposition to \reff{decomp} holds
on each side of the mirror
when the mirror travels from $\scrym$ to $\scryp$, as it is the case
for a uniformly accelerated system, see Sec. $2$.

\subsection{Partially transmitting mirrors}

In preparation for subsequent analysis,
we now present how to study partially transmitting mirrors.
In this case, one should also
consider simultaneously $U$ and $V$ modes.
Indeed, when the trajectory is asymptotically inertial, 
an incoming $U$ mode is partially scattered into an outgoing $V$ mode 
and partially transmitted as an outgoing $U$ mode.
Hence, when the mirror is not inertial, 
a proper description of the Bogoliubov coefficients requires to consider 
$2 \times 2$ matrices $\al,\bt$ which mix $U$ and $V$ modes. 

There are two different ways to describe partially transmitting mirrors.
First, one can choose from the outset the transmission
coefficient (expressed in the rest frame of the mirror) 
and deduce from it the Bogoliubov coefficients,
see Sec. II.B in \cite{OPa}.
We shall not follow this method since it does not allow 
to switch off the coupling to the radiation field. 

The other method is based on self-interactions described by an action.
The principle usefulness of this model is to allow 
to switch on and off the coupling of the radiation with the mirror.
We will see in the next sections that this is necessary to obtain 
well defined transition energy fluxes for a uniformly accelerated mirror.
In the following, we shall use only this model,
see Sec. III in \cite{OPa} for more details.

In this model, the scattering on the mirror is governed by an action
whose density is localized on the mirror trajectory $x^\mu_{cl}(\t)$,
where $\t$ is the proper time,
\ba \label{Hint}
L_{int} 
= - \int d\tau H_{int}(\t) = - g_0 \int d\tau g(\t) \int \! d^2x \: \delta^2(x^\mu-x^\mu_{cl}(\t)) \;
{\cal F}[\Phi^\dagger(t,z) \: ,\Phi(t,z) ] \ .
\ea
$g_0$ is the coupling constant.
The real function $g(\t)$ controls the time dependence of the interaction:
When the coupling lasts a proper time lapse equal to $2T$, 
$g(\t)$ is normalized by $\int d\t g(\t) = 2T$. 
To preserve the linearity of the scattering,
${\cal F}$ must be a quadratic form of the field $\Phi$
and to have a well-defined Hamiltonian, 
it should be hermitian. 
The various possibilities with the lowest number of derivatives
are ${\cal F}_0= \Phi^\dagger \Phi,
{\cal F}_1= \Phi^\dagger i \didi{\tau} \Phi$ and
${\cal F}_2= \di_{\tau} \Phi^\dagger \di_{\tau} \Phi$.

In the interacting picture,
the charged field evolves freely. It can thus be decomposed as
\ba \label{field}
\Phi(U,V) = \into \! \frac{d\om}{\sqrt{4\pi \om}} 
\left(
a^{U}_{\om} e^{-i\om U}
+ a^{V}_{\om} e^{-i\om V}
+ b^{U \: \dagger}_{\om} e^{i\om U}
+ b^{V \: \dagger}_{\om} e^{i\om V}
\right)  \ .
\ea   
The annihilation and creation operators 
of left and right-moving particles (and anti-particles)
are constant and obey the usual commutation relations 
\ba
[a^i_\om,a^{j \: \dagger}_{\omp}] = \delta^{ij} \delta(\om-\omp)
\; \; , \; \; 
[b^i_\om,b^{j \: \dagger}_{\omp}] = \delta^{ij} \delta(\om-\omp)
\ .
\ea
All other commutators vanish.
In the interacting picture, 
the states evolve through the action of a time-ordered operator 
$Te^{iL_{int}}$.
When the initial state is vacuum, 
up to second order in $g_0$,  the state on $\scryp$ is
\ba \label{evoluatedvac}
T e^{i L_{int}} \vac 
&=& \vac 
+ i L_{int}  \vac + \frac{(iL_{int} )^2}{2} \vac
+ \ket{D} \ .
\ea
The ket $\ket{D}$ contains 
terms arising from time-ordering. 
None of these terms contribute to the total energy emitted
(see \cite{OPa} for a detailed analysis).
Hence we drop $\ket{D}$ from now on. 

The relationship between this model
and the original Davies-Fulling model can be made explicit
by considering the case where
${\cal F}=\Phi^\dagger i \didi{\tau} \Phi$
and $g(\t) = 1$ (see \cite{OPa}).
In this case, whatever the mirror trajectory is,  
the first order transition amplitudes are related to
overlaps $\al_{\om\omp}^{ij},\bt_{\om\omp}^{ij}$
entering in \reff{decomp} in the following way 
\begin{subeqnarray} \label{Bogo}
A^{VV \; *}_{\om \omp} 
&\equiv&
\vacave{a^V_\om \:  e^{iL_{int}} \: a^{V \; \dagger}_{\omp}}_c
= \delta(\om - \omp) - i g_0  \, \al_{\om \omp}^{VV} \\
B^{VV \; *}_{\om \omp} 
&\equiv& 
\vacave{ e^{-iL_{int}} \: a^{V \: \dagger}_\om b^{V \: \dagger}_{\omp}}
= - i g_0 \, \bt_{\om \omp}^{VV} \\
A^{VU \; *}_{\om \omp} 
&\equiv&
\vacave{a^V_\om \: e^{iL_{int}} \: a^{U \; \dagger}_{\omp}}_c
= - i g_0 \, \al_{\om \omp}^{VU \; *} \\
B^{VU \; *}_{\om \omp} 
&\equiv&
\vacave{ e^{-iL_{int}} \: a^{V \: \dagger}_\om b^{U \: \dagger}_{\omp}}
= i g_0 \, \bt_{\om \omp}^{VU \; *} \ ,
\end{subeqnarray}
where the subscript $\langle \rangle_c$ means that 
only the connected graphs are kept.
In Eqs.(\ref{Bogo}b) and (\ref{Bogo}d),
one sees clearly the link between the $\bt$ coefficients
and pair creation amplitudes.

When using these amplitudes to compute energy fluxes
one encounters severe IR divergences 
due to the massless character of $\Phi$. 
These divergences can be eliminated
by considering the Hamiltonian 
with one more derivative
\ba \label{F}
{\cal F}[\Phi^\dagger , \Phi]
= \frac{dx^\mu_{cl}}{d\t} \frac{dx^\nu_{cl}}{d\t}
(\di_\mu \Phi^\dagger \di_\nu \Phi + \di_\mu \Phi \di_\nu \Phi^\dagger) \ .
\ea
The two terms within the parentheses mean that 
the interaction is symmetrical under charge conjugation.
This implies that the transition amplitudes will be invariant 
under the exchange of $a$ and $b$ operators. 

It should be stressed that the self-interacting model
can handle without modification in the cases when the 
trajectory enters and/or leaves space through null infinities.
In these cases, when computing perturbatively transition amplitudes,
one automatically adopts the convention of using asymptotically free modes. 
Indeed, the interacting picture is based on the assumption that 
the interaction is switched on and off asymptotically. This remarks
reinforces the well-founded character of the choice adopted
in the former sub-section to use, for the $out$-basis, free Minkowski modes
on $\scryp$ .

\subsection{Energy fluxes}

In this subsection, we compute physical observables 
such as the number of emitted particles, 
the energy and its fluxes, 
to second order in $g_0$
and when the initial state is vacuum.
We study only the left-moving quanta 
emitted toward $\scrypL$: 
all the results for the right-movers 
can be obtained 
by exchanging $U$ and $V$.

The mean number of $V$ particles of energy $\om$
is particularly simple
because only the second term of \reff{evoluatedvac} contributes.
One obtains
\be
\ave{N^V_\om} \label{N}
\equiv \vacave{e^{-iL_{int}} \: 
a^{V \: \dagger}_\om a^V_\om \: e^{iL_{int}}}
= \Io{\omp} 
\left( \left| B^{VV}_{\om \omp} \right|^2 
+ \left| B^{VU}_{\om \omp} \right|^2 \right)\, ,
\ee
in the place of \reff{NDF} since the partner of $  a^V_\om $ can be either 
a $U$ or a $V$ quantum. 
Then the (subtracted) integrated energy is, as usual,
\ba
\ave{H_M^V} &=& 2 \Io{\om} \om \ave{N^V_\om} \ , \label{H}
\ea
where the factor of $2$ stands for particles + antiparticles.

One can also compute the local flux of energy.
The corresponding Hermitian operator is
$\TVV = \di_V \Phi^\dagger \di_V \Phi + \di_V \Phi \di_V \Phi^\dagger$. 
Its expectation value is given by 
\ba
\ave{\TVV(V)} &\equiv& \label{TVV1}
\vacave{e^{-iL_{int}} \: \TVV \: e^{iL_{int}}}  - \vacave{\TVV}
\nonumber \\
&=& \label{TVV4} \ave{\TVVI} + \ave{\TVVII} \, , 
\ea
where 
\ba 
\label{TVVI}
\ave{\TVVI} &\equiv&
\vacave{L_{int} \TVV L_{int}}_c  \nonumber \\
&=&
 2 \sum_{j=U,V} \int \! \Io{\om} \! \! d\omp 
\frac{\sqrt{\om \omp}}{2 \pi} 
e^{-i(\omp-\om)V} 
\left( \Io{k} B^{Vj \: *}_{\om k} B^{Vj}_{\omp k} \right) \, ,
\ea
and 
\ba \label{TVVII}
\ave{\TVVII}
&\equiv& 
- 2 \left[ \;\im\left\{ \vacave{\TVV L_{int}} \right\}
+ \re\left\{ \vacave{\TVV L_{int}L_{int}}_c \right\} \right] \label{TVV2} \nonumber \\
&=&
- 2 \re \Bigg\{
\sum_{j=U,V} \int \! \Io{\om} \! \! d\omp 
\frac{\sqrt{\om \omp}}{2 \pi}  
e^{-i(\omp+\om)V} 
\left( \Io{k} A^{Vj \: *}_{\om k} B^{Vj}_{\omp k} \right)
\Bigg\} \ .
\ea
We have subtracted the average value of $\TVV$ in the vacuum 
in order to remove the zero point energy.
In \reff{TVV4}, we have introduced
$\ave{\TVVI}$ which determines the integrated (positive) energy 
$\ave{H_M^V}$ of \reff{H} and
$\ave{\TVVII}$ which integrates to $0$.
Note that the linear term in $g_0$ 
in the first line of \reff{TVV2} reappears in the second 
through the definition of the $A$ terms given by Eqs.(\ref{Bogo}).

Besides these expressions based on the amplitudes $A$ and $B$, 
one can also express $\ave{T_{VV}}$
by using the ``scattered'' Wightman 
function\footnote{This illustrates the fact that 
one does not need to choose an $out$-basis when computing 
expectation values of local operators with initial states prepared on $\scrym$.
Notice however that the subtraction in \reff{TVVW1} implicitly reintroduces
the notion of Minkowski vacuum on $\scryp$ since the {\em only} singularity
of $W_{vac}(U,V ; U',V')$ in \reff{Wvac} is the usual short distance one,
independently of the presence of the mirror on $\scryp$. 
Thus, since subtracting the vacuum contribution on $\scryp$
is equivalent to subtracting that of the $out$-modes,
the use of \reff{Wvac} implies that the mirror is no longer coupled 
to the radiation field at asymptotically late times.}.
This function is defined by 
\ba \label{W}
W(U,V ; U',V') 
= \vacave{ e^{-iL_{int}} \:  \Phi^\dagger(U,V) \: \Phi(U',V') \: e^{iL_{int}} }_c \ . 
\ea
To obtain the subtracted flux, 
one needs also the unperturbed Wightman function
evaluated in the vacuum
\ba 
W_{vac}(U,V ; U',V') 
&=& \vacave{ \Phi^\dagger(U,V) \: \Phi(U',V') } \nn
&=& - \inv{4\pi} \left( \ln ( V' - V - i\epsilon )  + \ln (U' - U - i\epsilon )\right) 
\label{Wvac} \ . 
\ea
In terms of these functions, the mean flux on $\scrypL$ reads \cite{OPa} 
\ba 
\ave{\TVV(V)}
&=& \label{TVVW1}
2 \lim_{V' \rightarrow V} \di_V \di_{V'}
\left[ W(U,V ; U',V') - W_{vac}(U,V ; U',V') \right] \ .
\ea
Notice finally that this expression also applies to
the Davies-Fulling model and leads to
the well-known result
\ba
\ave{\TVV(V)}^{DF}
&=& \label{TVVW2} 
-{1 \over 2\pi} \lim_{V' \to V}\lim_{\e\to 0}
\partial_{V} \partial_{V'} 
\left[ 
\ln ( U_{cl}(V') - U_{cl}(V) -i\e ) 
- \ln ( V' - V -i\e ) \right] \nonumber \\ 
&=& \label{TVVW3}
\frac{1}{6\pi}
\left\{ 
\left( \frac{dU_{cl}}{dV} \right)^{1/2}
\di_V^2 \left[ \left( \frac{dU_{cl}}{dV} \right)^{-1/2} \right] 
\right\}
\ .
\ea
From this equation one sees that the energy flux is local 
in that it contains at most three derivatives of $U_{cl}(V)$ 
evaluated at the advanced time $V$.
When considering the interacting model with $g(\tau)
\neq \mbox{const.}$, this local property will be lost.

\section{Uniformly accelerated mirrors}

Uniform acceleration means that
\ba
\frac{d^2x^\mu}{d\tau^2} \frac{d^2x_\mu}{d\tau^2}= -a^2 = \mbox{const.} 
\ea
In terms of Minkowski space-time coordinates, the trajectory reads
$t^2 - z^2 = U V = -1/a^2$.
In the sequel, we will consider a uniformly accelerated mirror 
living in the right Rindler wedge $R$,
\ie its trajectory is $U_{cl}(V)=-1/a^2V$ 
with $V$ running from $0$ to $+\infty$,
see Fig. $3$. 
\begin{figure}[ht] 
\epsfxsize=6.5cm
\psfrag{scrypU}{${\scrypR}$}
\psfrag{scrypV}{${\scrypL}$}
\psfrag{scrymU}{${\scrymR}$}
\psfrag{scrymV}{${\scrymL}$}
\psfrag{im}{$i^-$}
\psfrag{ip}{$i^+$}
\psfrag{U}{$U$}
\psfrag{V}{$V$}
\centerline{{\epsfbox{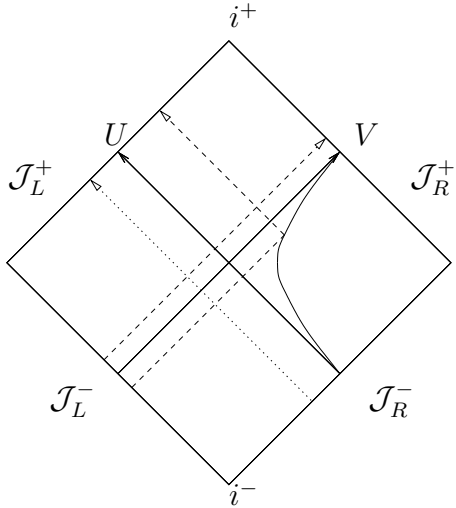}}}
\caption{In this Penrose diagram, 
we show a uniformly accelerated trajectory 
going from $V=0$ on $\scrymR$ to $U=0$ on $\scrypR$.
The dashed lines represent the scattering of
a pair of quanta (represented by localized wave-packets)
emerging from $\scrymL$.
One particle is reflected into an outgoing $V$ quantum for $U<0$
whereas the other ends as a left-mover on $\scrypR$ 
for $U>0$. The dotted line represents a $V$-quantum which is not
reflected by the mirror.}
\end{figure}

The scattering associated with this trajectory leads to several difficulties 
when using the Davies-Fulling model.
In Sec. $2.1$, we list these difficulties.
Then we will see how they can be resolved by using 
our self-interacting model 
with a smooth switching on and off coupling. 

\subsection{The difficulties using the Davies-Fulling model} 

When considering the scattering by a uniformly accelerated mirror
with the Davies-Fulling model, on the left side of the trajectory,
the image on $\scryp$ of the scattered $in$-modes are 
\ba \label{Umodescat}
\varphi_{\om}^{U,in} (U,V) 
&=&
\Theta(U) \frac{e^{-i{\om} U}}{\sqrt{4\pi {\om}}} 
+ \Theta(V) ( - \frac{e^{i{\om} /a^2V}}{\sqrt{4\pi {\om}}} ) \\
\varphi_{\om}^{V,in} (U,V) 
&=&
\Theta(-V) \frac{e^{-i{\om} V}}{\sqrt{4\pi {\om}}}  \ .
\ea
As expected from Sec. $1.2$, they are singular where the
mirror enters ($V=0$) and leaves ($U=0$) the space-time.
Their overlaps with plane waves ($out$-modes) are \cite{Grove1}
\begin{subeqnarray} \label{boguam}
\al_{{\omp} {\om}}^{VV \: *} 
&\equiv&
\left( \varphi_{{\omp}}^{V,out},\varphi_{{\om}}^{V,in} \right)
= \inv{4\pi} \frac{{\om}+{\omp}}{\sqrt{{\om}{\omp}}} 
\frac{i}{{\om} - {\omp}} \\
\bt_{{\omp} {\om}}^{VV \: *} 
&\equiv&
\left(\varphi_{{\omp}}^{V,out \: *},\varphi_{{\om}}^{V,in} \right)
= \inv{4\pi} \frac{{\om}-{\omp}}{\sqrt{{\om}{\omp}}} 
\frac{i}{{\om} + {\omp}} \\
\al_{{\omp} {\om}}^{VU \: *} 
&\equiv&
\left(\varphi_{{\omp}}^{V,out},\varphi_{{\om}}^{U,in}\right)
= \frac{-1}{\pi a} 
K_1(- i\frac{2 \sqrt{{\om} {\omp}}}{a}) \\
\bt_{{\omp} {\om}}^{VU \: *} 
&\equiv&
\left(\varphi_{{\omp}}^{V,out \: *},\varphi_{{\om}}^{U,in}\right)
= \frac{- i}{\pi a} 
K_1(\frac{2 \sqrt{{\om} {\omp}}}{a}) \ ,
\end{subeqnarray}
where $K_1(z)$ is a modified Bessel function 
(see Appendix A).

When computing $\ave{N_\om}$ 
(given by $\Io{\omp} ( |\bt^{VV}_{\omp\om}|^2 + |\bt^{VU}_{\omp\om}|^2 )$),
see Eqs.(\ref{NDF}) and (\ref{N})), 
the integral over $\om'$ diverges in the IR.
Moreover, $\al_{\omp \om}^{VV}$ diverges when $\om=\omp$.
Similarly, $\al_{\omp \om}^{VU}$ is ill-defined since the 
integral representation of the Bessel function in Eq.(\ref{boguam}c)
requires to contain a finite an positive real part,
see \reff{Bessel1}.

In addition to these problems in momentum space,
when computing the space-time properties of the flux, 
one encounters the following properties.
When plugging
$U_{cl}(V)=-1/a^2V$ for $V>0$ in \reff{TVVW3},
one finds that $\ave{\TVV(V)}^{DF}$ vanishes\footnote{Generally, 
the trajectories which provide
a vanishing $V$ flux are of the form 
$U_{cl}(V)=\frac{AV+B}{CV+D}$ \cite{DaviesFulling}. 
Time-likeness imposes $AD>BC$.
If $C=0$, we recover inertial trajectories.
If $C\neq 0$, we recover uniformly accelerated trajectories 
with $a=C/\sqrt{AD-BC}$.}.
This is not in agreement with the non-vanishing character of the $\beta$ since,
on one hand, $\ave{H_M^V} = \Io{\om} \om \ave{N_\om^V}$
and on the other hand, $\ave{H_M^V} = \Ie{V} \ave{T_{VV}}$.
It is as if the created particles were carrying no energy flux 
\cite{DaviesFulling,DaviesFulling3,Walker}.

To sum up, the difficulties of the Davies-Fulling model 
for uniformly accelerated trajectories are:
\begin{itemize}
\item unregulated and IR diverging overlaps, Eqs.(\ref{boguam}), 
\item a diverging expression for $\ave{N_\om}$, the mean number of particles created,
\item a vanishing local flux although pair-creation transition amplitudes 
do not vanish,
\item when one considers the scattering by {\it two} uniformly accelerated mirrors
with symmetrical trajectories, i.e. which both obey $UV= - 1/a^2$,  
the (unregulated) overlaps $\bt$ also vanish, together with
the local flux. From \cite{Gerlach}, one could infer that these settings  
form a perfect interferometer.
This cannot be exactly the case since the two mirrors live 
in two causally uncorrelated regions.
This issue will be fully discussed in a forthcoming paper \cite{OPa3}.
\end{itemize}

\subsection{The switching function $g$}

To avoid the difficulties listed above, 
we shall use our self-interacting model, 
based on Eqs.(\ref{Hint}) and (\ref{F}).
We require that the switching function $g(\t)$ be 
continuous, differentiable 
and that it decrease sufficiently rapidly for large $\t$.
A choice  we find very convenient and shall adopt is
\ba \label{g}
g(\t) = e^{-\disp 2\eta \disp \cosh(a\t)} 
= e^{-\disp \eta (aV_{cl}(\t)+1/aV_{cl}(\t))}
\ ,
\ea
where $0<\eta \ll 1$ is a dimensionless parameter 
which controls the switching time.

$g(\t)$ can be interpreted in two different ways:
either as governing how the interactions with the mirror are 
turned on and off, or as a mathematical regulator which properly defines
the transition amplitudes $A_{ij}^{\om\omp}$, $B_{ij}^{\om\omp}$.
This second interpretation implicitly relies on the limit $\eta \to 0$,
see  Appendix B. In the body of the text, we
shall use the first (physical) interpretation of $g(\t)$.

\begin{figure}[ht] \label{gtaufig}
\epsfxsize=6.5cm
\epsfysize=5.5cm
\psfrag{tau}{${\tau}$}
\centerline{\epsfbox{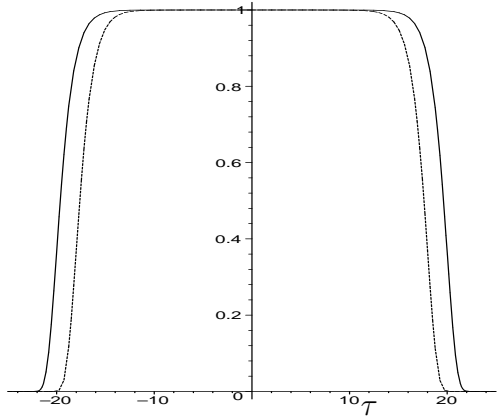}}
\caption{Here are presented two $g(\tau)$, both for $a=1$.
The solid line is with $\ln\eta=-20$ and the dashed line with $\ln\eta=-18$.
One sees that the size of the plateau is linear in $\ln\eta$ 
whereas the slope is independent of $\eta$.
This remark will be crucial when considering
the Rindler energy emitted by the mirror, see Sec. $3$.} 
\end{figure}

The features of $g(\t)$ are the following (see Fig. $4$) : 
\begin{description}
\item[a)] a plateau of height $1$ centered around $\t=0$
and of width equal to 
\ba 2T \equiv 2\pi g_{\la=0} \simeq \frac{2 | \ln(2\eta) |}{a} \ea
\item[b)] slopes which are maximal 
and equal to $a/e + {\cal O}(\eta^2)$
for $a\t \simeq \pm (aT + \ln2)$, 
\item[c)]  an exponentially decreasing tail.
We shall see that this extremely rapid decreasing behavior
($\sim e^{-\eta e^{a|\t|}} $)
is sufficient for having a finite Minkowski flux on $\scryp$.
\end{description}

We now compute the Rindler and Minkowski Fourier components of $g$ 
since they will reappear in the expressions of the transition amplitudes.
They are given in terms of the modified Bessel functions: 
\ba 
\label{gla}
g_\la 
&\equiv& \inv{2\pi} \Ie{\t} e^{- 2\eta \cosh(a\t)} e^{i\la\t} 
= \inv{a\pi} K_{i\la/a}(2\eta) \\
\label{gom}
g_\om 
&\equiv& \inv{2\pi} \Io{V} e^{-\eta(aV+1/aV)} e^{i\om V} 
= - \inv{a\pi} \inv{\sqrt{1-i\om/a\eta}} K_{1}(2\eta\sqrt{1-i\om/a\eta}) \ .
\ea
In the UV, for $\la\gg a$ and $\om \gg a/\eta$,
the Rindler and Minkowski components decrease respectively as
\ba 
|g_\la| &\stackrel{\la \rightarrow \infty}{\sim}& \label{UVl}
\inv{a} (a/\lambda)^{1/2} e^{-\pi \la / 2 a} \\
|g_\om| &\stackrel{\om \rightarrow \infty}{\sim}&  \label{UVo}
\inv{a} (a/\om)^{3/4} \eta^{1/4} e^{-\sqrt{2\om\eta/a}} \ .
\ea
From \reff{UVl}, one sees that 
the UV behavior of the Rindler component is independent of $\eta$. 
This is a direct consequence of point {\bf b)} which states 
that the maximal slope of the switching function is independent of $\eta$
when expressed in proper time $\t$. 
From  \reff{UVo} one finds instead that 
the UV behavior of the Minkowski components 
is damped by the regulator $\eta$.

\subsection{Regularized amplitudes and particle content}

Given $g(\t)$ of \reff{g}, the transition amplitudes 
can be explicitly calculated. They are given in the Appendix B 
and, as expected, they are well-defined.
Nevertheless, we will not work with these amplitudes 
characterized by Minkowski frequencies
since they are not convenient to compute
the expectation value of observables.
Similarly, we will not work with 
the transition amplitudes with Rindler frequencies
even though they are simply expressed in terms of $g_\la$ of  \reff{gla}. 

It turns out that it is more convenient to express the fluxes and the energy 
in terms of transition amplitudes containing one Minkowski and one Rindler 
frequency.
More precisely, these amplitudes mix Minkowski and ``Unruh'' quanta. 
The Unruh modes $\hat \varphi_\la^j$
are linear combinations of positive frequency Minkowski modes
{\underline {and}} eigenmodes of Rindler energy $\la$ 
(see \cite{Unruh,PhysRep} and Appendix C). 

These ``mixed'' transition 
amplitudes\footnote{These scattering amplitudes make contact
with the transition amplitudes of a uniformly accelerated detector 
coupled to the radiation field $\Phi$ \cite{Unruh,Matsas,MaPa,PhysRep,UW}.
For instance, the spontaneous emission amplitude of a two-level atom 
with an energy gap $\Delta M$
is equal to $(\om\sqrt{\la})^{-1}B^{VU}_{\om,\la}|_{\la=\Delta M}$,
see Eq.(2.48) in \cite{PhysRep}.} are given 
by the matrix elements of the scattering operator 
with the Minkowski operator $a^j_\om$ 
and the Unruh one $\a_\la^j$:
\ba
A^{ij \; *}_{\om \la} 
&\equiv& \vacave{a^i_\om \:  e^{iL_{int}} \: \a^{j \; \dagger}_{\la}}_c \nonumber \\
B_{\om \la}^{ij}
&\equiv& \vacave{a_\om^i \: \b_\la^j \: e^{iL_{int}}} \ .
\ea 
To first order in $g_0$, 
using Eqs.(\ref{Hint}),(\ref{F}),(\ref{Bessel1}) and (\ref{78}),
we get
\ba 
A^{VU \: *}_{\om \la} \label{regA}
&=& - \frac{ig_0}{\pi a} \; \sqrt{\frac{\om\la}{1-e^{-2\pi\la/a}}} \;
(1-i\om/a\eta)^{-(1-i\la/a)/2} \; 
K_{1-i\la/a}(2\eta\sqrt{1-i\om/a\eta}) \\
B^{VU}_{\om \la} \label{regB}
&=& \frac{ig_0}{\pi a} \; \sqrt{\frac{\om\la}{1-e^{-2\pi\la/a}}} \;
(1-i\om/a\eta)^{-(1+i\la/a)/2} \; 
K_{1+i\la/a}(2\eta\sqrt{1-i\om/a\eta}) \ .
\ea 
Moreover, as shown in the Appendix C, 
$U$ and $V$ Unruh modes coincide when evaluated along the trajectory.
Therefore $B_{\om \la}^{VU}=B_{\om \la}^{VV}$ 
and similarly for $A$.

Using these amplitudes, the mean Minkowski energy emitted 
to $\scrypL$ can be written
\ba  \label{Hunruh}
\ave{H_M^V} = 4 \Io{\om} \om \Ie{\la} \left| B_{\om \la}^{VV} \right|^2 
= \Io{\om} \Ie{\la} h_M(\om, \la; \eta)
\ .
\ea
The number $4$ before the first integral
means that processes involving $U$ or $V$ particles and $U$ or $V$ anti-particles
equally contribute to the mean Minkowski energy emitted to $\scrypL$
(or $\scrypR$).

The exact computation of $\ave{H_M^V}$ will be performed in the next section. 
In the following, we shall study $h_M(\om,\la,\eta)= 4
\om \left| B_{\om \la}^{VV} \right|^2 $ 
to give a qualitative understanding of  $\ave{H_M^V}$. 
The usefulness of the mixed representation is that 
the behavior of $h_M(\om, \la; \eta)$ in the $(\om,\la,\eta)$ space
is quite easy to explain.
The starting point is that the norm of $B_{\om \la}^{VU}$ decreases as 
$e^{-2\pi|\la|/a}$ for $|\la| \gg a$,
as expected from the correspondence mentioned in the last footnote.
Hence, the relevant range of $\la$ is centered around $0$ and of extension a few $a$'s.
When $\la$ belongs to this interval, the following analysis applies.

First, $\eta$ acts as a regulator: 
the Minkowski frequencies which contribute to $h_M$ belong to the 
interval
\ba \label{omegarange} a\eta \lesssim \om \lesssim \xi a/\eta \ , \ea  
where $\xi$ is a numerical factor.
Its value is $\sim 0.50$
when one uses the mid height criterion:
$h_M(\om=\xi a/\eta,\lambda;\eta) / h_M(\om=a,\lambda;\eta) = 1/2$
with $\la$ belonging to the relevant interval.

Secondly, within the range given in \reff{omegarange}, 
$h_M$ hardly depends on $\om$, as shown in 
Fig.$5$.\footnote{In this, we recover what was found for
an accelerated detector, see Sec. $2.4$ of \cite{PhysRep}.
In that case, the Minkowski frequencies which contribute to the
processes are the Doppler frequencies which resonate with the energy gap 
$\Delta M$ when the interactions are turned on, 
i.e. those which satisfy $\om = e^{a \t}\Delta M$ for $-T<\t<T$.
Moreover, within that range, the transition amplitudes do not depend
on $\om$ and the limit $\eta\to 0$ can be used to estimate the amplitude,
as in \reff{Bgamma}.}
Hence, for any given value of $\eta$, 
one can first trivially perform the integral over $\om$ 
from $a\eta$ to $\xi a/\eta$. 
The value of this integral is given by $h_M(\om=a,\la;\eta) \times \xi a/\eta$,
since $\eta \ll 1$.
\begin{figure}[ht] \label{wBR2_lneta_lnomega}
\epsfxsize=6.5cm
\epsfysize=5.5cm
\psfrag{lnomega}{${\ln\om}$}
\psfrag{lneta}{${\ln\eta}$}
\centerline{\epsfbox{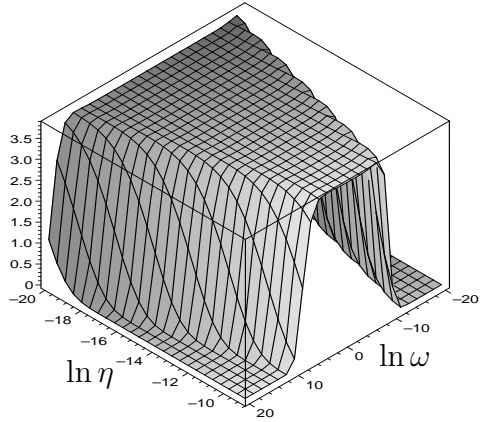}}
\caption{Here we show $h_M(\om,\la=0.15;\eta)$ 
in terms of $\ln\om$ and $\ln\eta$. 
One sees clearly that the surface exhibits a ``plateau'' of constant height which is 
limited by the lines $\ln\om = \pm \ln\eta$ . 
$h_M(\om,\la;\eta)$ is given in arbitrary units and $a=1$.} 
\end{figure}

Thirdly, the height of the plateau  hardly depends on $\eta$, see Fig.$6$.
\begin{figure}[ht] \label{wBR2_lambda_lneta}
\epsfxsize=6.5cm
\epsfysize=5.5cm
\psfrag{lambda}{${\la}$}
\psfrag{lneta}{${\ln\eta}$}
\centerline{\epsfbox{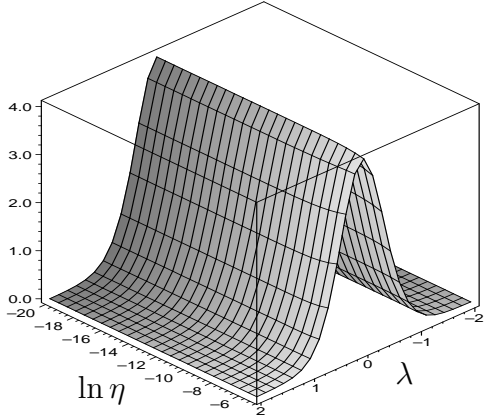}}
\caption{Here we show $h_M(\om=1,\la;\eta)$ 
in terms of $\lambda$ and $\ln\eta$. 
One clearly sees that the energy density $h_M$ 
is independent of $\eta$. 
This comes from that the fact that we compute it ``within the plateau'',
\ie for $a\eta \ll \om=a \ll a/\eta$. 
Again, $h_M(\om,\la;\eta)$ is given in arbitrary values and $a=1$.} 
\end{figure}
This can be understood from \reff{limtozero}: when \reff{omegarange} is satisfied
and when $\eta \ll 1$, $B_{\om \la}^{VU}$ is independent of $\eta$.
Hence, one can take the limit $\eta \to 0$ to estimate how $h_M$ depends on 
$\la$.
In this limit, the scattering amplitudes obey
\ba  \label{Bgamma}
B^{VV}_{\om \la} 
&\stackrel{\eta\rightarrow 0}{\longrightarrow}&
-i g_0 \frac{\la}{e^{\pi\la/a} - e^{-\pi\la/a}} 
\; \gamma_{\la \om}^{V \: *} \, ,
\ea
where $\gamma^V_{\la\om} = ({\varphi_\om^V},{\hat \varphi_\la^V})$
is given in \reff{gamma2} of Appendix C.
In this case, we get
\ba
h_M(\om=a,\la;\eta) &=& 4 \om \left| B_{\om \la}^{VV} \right|^2 \nn
&=&  \frac{g_0^2}{2\pi a} \left(\frac{\la}{e^{\pi\la/a} - e^{-\pi\la/a}}\right)^2 \ ,
\ea
thereby recovering the above mentioned Boltzmann expansion law 
for $|\la| \gg a$.
Thus the mean energy can be approximated by
\ba
\ave{H_M^V}
&=& \xi \frac{a}{\eta} \Ie{\la} h_M(\om=a,\la;\eta)  \nn
&=& \xi \frac{g_0^2 a^3}{6\pi\eta} \label{HH} \ .
\ea
Although the numerical factor is not exactly determined
(because of the ambiguity of defining $\xi$), 
we will see in the next section 
that the factor $g_0^2 a^3$ 
and the scaling in $1 / \eta$
correctly define the behavior of $\ave{H_M^V}$ 
when $\eta \ll 1$ (The exact value of $\xi$ is $3/8$ instead of $1/2$.).

So far, by the introduction of the switching off function (\reff{g}),
we have solved the first two problems listed in Sec. $2.1$.
Indeed, the transition amplitudes 
(expressed in the Minkowski, Rindler or mixed representation)
are now well-defined both in the IR and the UV, 
and the energy emitted is no more infinite.
The third and last point, \ie the issue concerning the local fluxes,
is the subject of the next section.

\section{Local Minkowski and Rindler fluxes}

In the Davies-Fulling model, the flux is given by \reff{TVVW3}.
It is a local function of the trajectory $U_{cl}(V)$
and its derivatives expressed at the advanced time $V$.
This result relies on the time independence of the coupling.
Indeed, this feature no longer occurs
when the coupling to the radiation field is 
switched on and off.

To compute $\ave{\TVV}$ of \reff{TVV1}
we first put together the two terms quadratic in $g_0$.
This is appropriate when computing local properties in space-time because
it leads to a simplification since this gives rise to a commutator which is local.
Then the flux reads
\ba \label{TVVcom}
\ave{\TVV(V)} 
&=& - 2 \im\left\{ \vacave{\TVV L_{int}} \right\} 
+ \re(\vacave{L_{int}[\TVV,L_{int}]} ) \\
&=& \ave{\TVV(V)}_1 + \ave{\TVV(V)}_2 \label{TVVcom2}\ .
\ea
(Note that all disconnected diagrams automatically cancel in this expression.)
Both terms are governed by the second derivative of the Wightman function, \reff{Wvac},  
\ba
\di_V \di_{V'} W_{vac}(V-V') 
=- \inv{4\pi} \inv{(V-V'-i\e)^2} \ .
\ea
Using this function, they read
\ba
\ave{\TVV(V)}_1 &=& \label{TVV1g}
- \frac{g_0}{2\pi^2} \im \left\{ 
\Ie{\t} g(\t) 
\dot V_{cl}^2(\t)
\inv{(V-V_{cl}(\t)-i\e)^4} 
\right\} \\
\ave{\TVV(V)}_2 &=& 
\frac{g_0^2}{2\pi^2} \re \Bigg\{ 
\Ie{\t'} \frac{g(\t')}{(V_{cl}(\t')-V-i\e)^2} 
\Ie{\t} g(\t) \:
\di_V \delta(V_{cl}(\t)-V)  \label{TVV2g} \\
&& \left( \frac{i}{(V_{cl}(\t')-V_{cl}(\t)-i\e)^2} +
\frac{\dot U_{cl}(\t) \dot U_{cl}(\t')}{\dot V_{cl}(\t) \dot V_{cl}(\t')} 
\frac{i}{(U_{cl}(\t')-U_{cl}(\t)-i\e)^2}
\right)
\Bigg\} \ , \nonumber
\ea
where a dot designates a derivation with respect to the proper time.
The function $\di_V \delta$ comes from the commutator: 
\ba
[\di_V\Phi^\dagger,\di_{V'}\Phi] 
= [\di_V\Phi,\di_{V'}\Phi^\dagger]
= \frac{i}{2} \partial_V \delta(V-V') \ .
\ea

To evaluate the integrals, it is appropriate 
to use the dummy variable $\tilde V = V_{cl}(\t)$ 
and to define a new function 
\ba \label{G}
G(\tilde V) = \dot V_{cl}(\t[\tilde V]) \: g(\t[\tilde V])\, , 
\ea
which can be interpreted as the effective coupling constant 
when using $\tilde V$ as the time.
Using this function, one can evaluate Eqs.(\ref{TVV1g}) and (\ref{TVV2g})
by integrating by part until the exponent of the pole is unity.
All boundary contributions vanish if $g(\t)$ decreases faster than $e^{- a|\t|}$,
a condition satisfied by the switching function 
we chose in \reff{g}. 
If  $g(\t)$ decreases slower than $e^{-a|\t|}$, 
the expectation value of $T_{VV}$ is ill-defined. 
Hence it appears that the condition $g(\t) e^{a|\t|} \to 0$ for $\t \to \pm \infty$ 
is a necessary condition for having well-defined Minkowski expressions.

Concerning \reff{TVV1g}, 
after three integration by parts, the last integration is trivially performed by using 
$\im \{ (x - i\e)^{-1} \} = \pi \delta(x)$. 
Concerning \reff{TVV2g}, the two terms within the parentheses are 
equal.
In order to compute this expression,
one first perform the integral over $\t$ by using the
function $\di_V \delta$. 
Then, as for \reff{TVV1g}, one integrates by parts until one gets a first order pole.

Grouping the results for Eqs.(\ref{TVV1g}) and (\ref{TVV2g}), one obtains, for $V>0$,
\ba
\ave{\TVV(V>0)}  \label{TVVG1}
&=& \inv{12 \pi} 
\left(
g_0 \di_V^3G 
- g_0^2 (G \di_V^4G+2\di_VG\di_V^3G)
\right) \\
&=& \frac{g_0^2}{12 \pi} \label{TVVG2}
\left( (\di_V^2 G)^2 + \di_V [...] \right) \ .
\ea
For $V<0$, one gets $\ave{\TVV} \equiv 0$ as expected
since the $V<0$ part of ${\scrypL}$ 
is causally disconnected to the mirror trajectory\footnote{We would
like to briefly comment on causality.
When computing the flux in a causally disconnected point 
with respect to the trajectory, one must obviously find zero
\cite{UW}. 
This is trivially the case when one expresses,
in  Eqs.(\ref{TVVI}) and (\ref{TVVII}), the transition amplitudes 
as integrals over the proper time $\t$
and first performs the integral over $\om$. 
However,  in the absence of a regulator,
one looses causality when inverting the order of the integrations.
This can be seen from the unregulated amplitudes where the fact that
mirror was in the $R$ or $L$ quadrant is lost, 
see Eqs.(\ref{unregABM}) in Appendix $B$.
The advantage of $g(\t)$ is to give rise to transition amplitudes,
see Eqs.(\ref{regABM}), wherein the prescription of the pole governed by $\eta$ 
keeps control on causality.
The same remark applies to the amplitudes in the mixed representation
defined in Eqs.(\ref{regA}) and (\ref{regB}). 
When using them to compute the flux, causality is kept.
Because of the analogy with the transition amplitudes
of an accelerated detector, causality is also preserved by regularizing these amplitudes,
see Eqs.(25),(26) and (63) of \cite{MaPa}.}.

In \reff{TVVG2}, we have separated the flux into two parts,
a square term which will lead to a positive Minkowski energy
$\ave{H_M^V} = \Ie{V} \ave{\TVV}$
and a total derivative which does not contribute to it
(note the similarity with $\ave{T_{VV}^I}$ and $\ave{T_{VV}^{II}}$ of \reff{TVV4}).
When using the coupling function of \reff{g} and Eqs.(\ref{TVVG2}) and (\ref{Bessel1}),  
one can obtain an analytical expression for $\ave{H_M^V}$
\ba
\ave{H_M^V} \label{HM}
&=& \frac{g_0^2 a^3}{6\pi} 
\left(
8 \eta^3 K_0(4\eta)  
+(4\eta^2+2\eta^4) K_1(4\eta) \right. \nn
&& \; \; \left.
-8\eta^3 K_2(4\eta)
-3\eta^4 K_3(4\eta)
+\eta^4 K_5(4\eta)
\right)
\ea
When taking the limit $\eta \to 0$ one obtains
\be
\ave{H_M^V} \label{HM0}
\stackrel{\eta \rightarrow 0}{\sim}
\frac{g_0^2 a^3}{16\pi} \inv{\eta} + {\cal O}(1) \ .
\ee
Hence, up to a numerical factor, 
we recover the result of \reff{HH}.
It is interesting to see how 
the pathological features of constant coupling 
re-emerge when taking $\eta \longrightarrow 0$.
In this limit, the effective coupling constant of \reff{G} obeys $G(V)=aV$.
Hence \reff{TVVG1} gives a vanishing flux 
whereas $\ave{H_M^V}$ clearly diverges (see \reff{HM0}).

To complete the analysis of the transients,
we now compute the Rindler flux ($\ave{T_{vv}} \equiv (dV/dv)^2 \ave{T_{VV})}$)
in terms of the Rindler advanced time $v=\inv{a} \ln(aV)$.
This analysis clearly establish that the Rindler energy carried by
the transients effects is insensitive to the duration of the interaction and
only depends on the rate of switching on and off the interactions. 
From \reff{TVVG1}, we get 
\ba
\ave{T_{vv}(v)} \label{Tvvg1}
&=& \frac{a^2}{12\pi} \Bigg[
- g_0 \di_v g + g_0^2 \left( g \di_v^2 g + 2 (\di_v g)^2 \right) 
\Bigg] \nn
&& - \inv{12\pi} \Bigg[
- g_0 \di_v^3 g + g_0^2 \left( g \di_v^4 g + 2 \di_v g \di_v^3 g \right) 
\Bigg] \\
&=& \frac{g_0^2}{12\pi} 
\left( a^2 (\di_v g )^2 + (\di_v^2 g )^2 \right) + \di_v[...] 
\ . \label{Tvvg2}
\ea
\begin{figure}[ht] 
\epsfxsize=6.5cm
\epsfysize=5.5cm
\psfrag{v}{$v$}
\centerline{\epsfbox{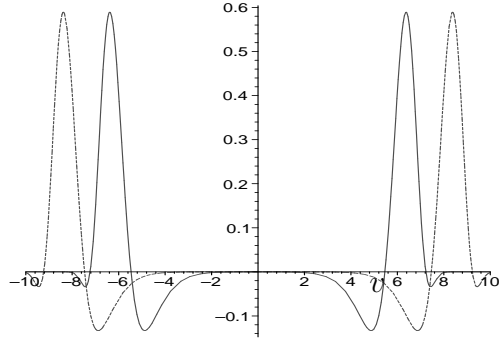}}
\caption{Here is plotted $\ave{T_{vv}(v)}$ 
as a function of $av=\ln(aV)$, for $a=1$ in arbitrary units
and for two different values of $\eta$.
The switching function has been taken for $\ln\eta=-6$ (plain curve) 
and $\ln\eta=-8$ (dashed curve). 
One can notice the two following properties.
The flux is significantly non-zero within the transients only.
The amplitude of these transients is independent of duration of the 
interactions governed by $\eta$.} 
\end{figure}
As for Minkowski energy $\ave{H_M^V}$, 
\reff{Tvvg2} shows that one obtains also a positive Rindler energy
$\ave{H_R^V} = \Ie{v} \ave{T_{vv}}$.
Using $g$ given by \reff{g}, the Rindler energy is 
\ba
\ave{H_R^V} \label{HR}
&=& \frac{g_0^2 a^3}{2\pi} 
\eta^2 K_2(4\eta) \\
\ave{H_R^V} \label{HR0}
&\stackrel{\eta \rightarrow 0}{\longrightarrow}& 
\frac{g_0^2 a^3}{16\pi} + {\cal O}(\eta) \ .
\ea

In Fig.$7$, we have plotted the Rindler flux 
when the switching function $g$ is given by \reff{g}.
We previously noticed that the slope of $g(\t)$ does not depend on $\eta$. 
As the slope determines the Fourier content of $g_\la$,
one understands why we obtain a non-vanishing Rindler energy 
even in the limit $\eta\rightarrow 0$.
We could have chosen different switching functions $g(\t)$ such that their slope
would tend to zero.
In the limit, we would have found  $\ave{H_R}=0$, as for a constant coupling.
However, in this case, we would have necessarily obtained
a diverging Minkowski energy $\ave{H_M^V}$, since the
condition $g(\t) e^{a|\t|} \to 0$ for $\t \to \pm \infty$ would not have been fulfilled.
Hence $\ave{H_R}$ cannot be sent to $0$ if one requires to have a finite $\ave{H_M^V}$.

\section*{Conclusions}

By considering the self-interacting model 
defined by \reff{Hint}, with ${\cal F}$ given in \reff{F}
and $g(\t)$ given in \reff{g}, 
we have solved all the difficulties listed in Sec. $2.1$:
The transmission amplitudes are well defined and given in 
Eqs.(\ref{regA}), (\ref{regB}) and (\ref{regABM}),
the mean energy flux is given in \reff{HM}
and the local flux in \reff{TVVG1}. 
All these quantities are regularized by the parameter $\eta$
which controls the switching on and off of the coupling 
through  \reff{g}.

The important lesson which emerges from this analysis is the following: 
when expressing $\ave{\TVV}$ in terms of $A$ and $B$ as in
Eqs.(\ref{TVVI}) and (\ref{TVVII}),
the regulator $\eta$ should be sent to $0$ {\em after}
having integrated over $k$, $\om$ and $\omp$. 
In this, we recover what was found in \cite{Parentani}
when evaluating the energy density in the Rindler vacuum.
If instead one first sends  $\eta \to 0$, the unregulated expressions of the
scattering amplitudes are so poorly defined that
one even looses causality and crossing symmetry, see Appendix B.
Therefore, in the presence of horizons, or more generally
when the mirror enters or leaves space-time through null infinities,
it is mandatory to consider the scattering amplitudes
as distributions and not only as functions of frequencies belonging to $[0, +\infty[$.

In addition, by expressing the scattering amplitudes in the mixed representation,
see Eqs.(\ref{regA}) and (\ref{regB}), 
we have made contact with the physics of a uniformly accelerated
detector. Indeed, its absorption/emission transition amplitudes are given by the same
functions as the scattering amplitudes in the mixed representation. 
This strict correspondence establishes that the physics of uniformly accelerated
systems is dominated by the kinematics, namely (near) stationarity with respect to
proper (Rindler) time and (near) singular behavior due the exponentially growing 
blue-shift effects associated with uniform acceleration.

\appendix
\section*{Appendix A : Bessel functions}
In this appendix we recall some features of the modified Bessel functions
$K_\nu(z)$, where $(\nu , z) \in {\mathcal C}$ (see \cite{Abra} $p.374$). They 
can be expressed by the following integral representation 
\ba
K_\nu(z) 
&=& \label{Bessel1}
\Io{t} e^{-z \cosh(t)} \cosh(\nu t), 
\ea
where $ |\arg(z)| < \pi/2$. For 
$k \in {\mathcal N}$ and $\nu \in {\mathcal R}$, one has
\ba
\left(
\inv{z}\frac{\di}{\di z}
\right)^k
\left\{
z^\nu  K_\nu(z)
\right\}
&=&
e^{-i\pi k}z^{\nu-k}  K_{\nu-k}(z) \ , \\
\left(
\inv{z}\frac{\di}{\di z}
\right)^k
\left\{
z^{-\nu}  K_\nu(z)
\right\}
&=&
e^{i\pi k}z^{-\nu-k}  K_{\nu+k}(z) \ .
\ea
We also recall the asymptotic behavior 
of the $K$'s
for small and large arguments 
\ba\label{limtozero}
K_0(z) 
\stackrel{z \rightarrow 0}{\sim} -\ln(z) 
\; &\and& \; 
K_{\nu}(z)
\stackrel{z \rightarrow 0}{\sim} \frac{\Gamma(\nu)}{2}  (\frac{2}{z})^{\nu},
\; \; \; \for \Re(\nu)>0 
\ea
whereas 
\ba
K_{\nu}(z)
&\stackrel{z \rightarrow +\infty}{\sim}& \sqrt{\frac{\pi}{2z}} e^{-z} ,
\;\; \mbox{for all $\nu$} \ .
\ea

\section*{Appendix B : Regularized transition amplitudes}

The aim of this appendix is to give the exact
expressions of the regularized transition amplitudes
in terms of Minkowski frequencies. 
The main virtue of the regulator $\eta$
is to define them without ambiguity.
The direct evaluation of Eqs.(\ref{Bogo}) 
with $g(\t)$ defined by \reff{g} gives
\begin{subeqnarray} \label{regABM}
A^{VV \; *}_{\om \omp} 
&=& \delta(\om-\omp) 
- \frac{4 i g_0}{\pi} \frac{\sqrt{\vert\om \omp\vert}}{a} \frac{\eta^2}{X^2} K_2(X) \\
&& \where 
X=\disp 2 \eta \sqrt{1 - i (\om - \omp)/a\eta } \ , \nn
B^{VV}_{\om \omp} 
&=& \frac{4 i g_0}{\pi} \frac{\sqrt{\vert\om \omp\vert}}{a} \frac{\eta^2}{X'^2} K_2(X') \\
&& \where 
X'=\disp 2 \eta \sqrt{1 - i (\om + \omp)/a\eta } \ , \nn
A^{VU \; *}_{\om \omp} 
&=& - \frac{i g_0}{\pi} \frac{\sqrt{\vert\om \omp\vert}}{a} K_0(Y)\\
&& \where 
Y=2\sqrt{(\om/a +i \eta)(- \omp/a - i \eta)}
\ , \nn
B^{VU}_{\om \omp} 
&=& \frac{i g_0}{\pi} \frac{\sqrt{\vert\om \omp\vert}}{a} K_0(Y') \\
&& \where 
Y'= 2 \sqrt{(\om/a + i \eta)(\omp/a - i \eta)} \ . \nonumber
\end{subeqnarray}
Two important remarks should be made.
Because these amplitudes have been regularized, they possess analytical properties
which guarantee that they obey
crossing symmetry, that is
\ba A^{ij \; *}_{\om \omp} = - B^{ij}_{\om,\omp e^{i\pi}} \ ,\ea
see \cite{PaMa} for exploiting this symmetry in studying accelerated detectors.
Secondly,  had the mirror followed the accelerated trajectory 
in the left quadrant rather than in the right one, the corresponding transition amplitudes 
would have been obtained by simply replacing $\eta$ by $-\eta$.

We wish also to stress that $g(\t)$ can be considered as 
a mathematical regulator which properly defines the transition amplitudes.
Consider for instance $B^{VU}_{\om \omp}$. 
Using Eq.(\ref{g}), it is given by  
\ba
B^{VU}_{\om \omp} 
\propto \frac{\sqrt{\om\omp}}{a}  
\Io{V} \inv{V} e^{\disp i \left( 
(\om+i\eta)V - (\omp-i\eta)/a^2V
\right)} \ .
\ea
One clearly sees that the integral is now well defined both for $V \to 0$ and $V \to \infty$.

Finally, it is interesting to take the limit $\eta \to 0 $ 
to see how one recovers the singular amplitudes that one would have obtained
with a constant coupling. 
Using Eqs.(\ref{regABM}), in the limit $\eta\rightarrow 0$, we get
\begin{subeqnarray} \label{unregABM}
A^{VV \; *}_{\om \omp} 
&\longrightarrow&
\delta(\om-\omp) 
+ \frac{i g_0}{2\pi} \; a \; \frac{\sqrt{\vert\om \omp\vert}}{(\om-\omp)^2} \\
B^{VV}_{\om \omp} 
&\longrightarrow&
- \frac{i g_0}{2\pi} \; a \; \frac{\sqrt{\vert\om \omp\vert}}{(\om+\omp)^2} \\
A^{VU \; *}_{\om \omp} 
&\longrightarrow&
- \frac{i g_0}{\pi} \; 
\frac{\sqrt{\vert\om \omp\vert}}{a} \; 
K_0(-2 i  \frac{\sqrt{\vert\om \omp\vert}}{a})\\
B^{VU}_{\om \omp} 
&\longrightarrow&
\frac{i g_0}{\pi} \; 
\frac{\sqrt{\vert\om \omp\vert}}{a} \; 
K_0(2 \frac{\sqrt{\vert\om \omp\vert}}{a}) \ .
\end{subeqnarray}
Although the choice of our Lagrangian, based on \reff{F}, 
has removed the IR divergences, 
the $A$ terms are clearly ill-defined, as in the original Davies-Fulling model.
More importantly,  crossing symmetry and causality are both lost if one uses
these unregulated amplitudes. This clearly establishes that the 
Bogoliubov coefficients must be conceived as {\it distributions},
or at least analytical functions of $\om$ and $ \om'$, 
and not merely as functions defined from $[0, +\infty[$.
Thus the limit $\eta \to 0$ should be performed
only at the end of the calculation, after having performed all integrations.
This is because the limit  $\eta \to 0$ in general does not commute with these
integrations. 

\section*{Appendix C : The Unruh modes}

By definition, the ``Unruh" \cite{Unruh} modes 
$\hat \varphi_\la^V$ and $\hat \varphi_\la^U$ 
possess the following properties:
\begin{itemize}
\item they are made of positive Minkowski frequency modes only,
whatever the sign of $\la$ be,
\item they are eigenfunctions of $iaV\di_V$ (or $-iaU\di_U$)
with eigenvalue $\la$ . 
\end{itemize}
They are thus well-adapted to study uniformly accelerated systems 
since they are eigenmodes of 
$i \di_\t = \la$ where $\t$ is the proper time calculated along the accelerated trajectory.

Since the Unruh modes form a complete and orthonormal basis, 
one can define in a canonical way 
the corresponding annihilation and creation operators
of particles and anti-particles
$\a_\la^i, \a_\la^{i \: \dagger}, \b_\la^i$ and $\b_\la^{i \: \dagger}$: 
\ba
\a_{\la}^{i} 
= \left( \hat \varphi_\la^{i} , \Phi \right) \; , \;
\b_{\la}^{i} 
= \left( \hat \varphi_\la^{i} , \Phi^\dagger \right) \ ,
\ea
where the subscript ${}^i$ stands as before for $U$ and $V$.
Hence, the scalar field can be decomposed as
\ba
\Phi(U,V) = \sum_{i=U,V} \Ie{\la} 
\left(
\hat \varphi^{i}_\la \a_\la^i + \hat \varphi^{i \: *}_\la \b_\la^{i \: \dagger} 
\right) \ .
\ea
Note that the integrals over $\la$ cover the entire real axis.

The Unruh modes are analytically expressed by the following expressions 
\ba
\hat \varphi_\la^V(V) 
&\equiv& \label{VU2U} \lim_{\e\rightarrow 0}
\frac{[a(V-i\e)]^{-i\la/a}}{\sqrt{4\pi \la (1-e^{-2\pi\la/a})}} \\
&=&\label{VU2M}
\Io{\om} \gamma^V_{\la\om}
\frac{e^{-i\om V}}{\sqrt{4\pi\om}} 
\ ,
\ea
with 
\ba 
\gamma^V_{\la\om} 
&\equiv&\label{gamma1} ({\varphi_\om^V},{\hat \varphi_\la^V}) \\
&=& \label{gamma2}
\frac{\Gamma(-i\la/a)}{\sqrt{\frac{a \pi}{\la \sinh(\pi\la/a)}}}
{(\frac{\om}{a})}^{i\la/a} \frac{e^{-\om \e}}{\sqrt{2\pi a\om}} 
= \left[ {(\gamma^V)}^{-1}_{\la\om} \right]^* \ .
\ea
Notice that the regulator $\e$ in Eqs.(\ref{VU2U}) and (\ref{gamma2})
plays a role similar to $\eta$ in the text : 
$\e$ is {\em needed} to properly define the energy density in the Rindler vacuum \cite{Parentani}. 

When considering $U$ modes, we get similar expressions with
$\gamma^U_{\la\om} = {\gamma^V_{\la\om}}^*$.
Finally, when evaluated along the accelerated trajectories, 
within the right (R) or the left (L) quadrant,
$U$ and $V$ Unruh modes coincide and are given by
\ba
&\mbox{in}& R \; : \; \left\{
\begin{array}{l}
V = V_{cl}^R(\t) = e^{a\t}/a \\
U = U_{cl}^R(\t) = - e^{-a\t}/a
\end{array}
\right. \and
\hat \varphi_\la^V(V) = \hat \varphi_\la^U(U)
= \label{78}
\frac{e^{-i\la\t}}{\sqrt{4\pi \la (1-e^{-2\pi\la/a})}} \ , \\
&\mbox{in}& L \; : \; \left\{
\begin{array}{l}
V = V_{cl}^L(\t) = - e^{-a\t}/a \\
U = U_{cl}^L(\t) = e^{a\t}/a
\end{array}
\right. \and 
\hat \varphi_\la^V(V) = \hat \varphi_\la^U(U)
= 
\frac{e^{i\la\t}}{\sqrt{4\pi \la (e^{2\pi\la/a}-1)}} \ .
\ea


\end{document}